\documentstyle[12pt]{article}
\input epsf

\begin{document}


\newcommand{\be}{\begin{equation}}
\newcommand{\ee}{\end{equation}}
\def\bq{\begin{eqnarray}}
\def\eq{\end{eqnarray}}
\def\n{\nonumber}
\def\t{\tau}
\def\ti{\tilde}
\def\ii{\int_{-\infty}^{+\infty}}
\def\ep{\epsilon}
\def\de{\delta}
\def\a{\alpha}
\def\th{\theta}
\def\G{\Gamma}
\def\ph{\phi}
\def\s{\sigma}
\def\vp{\varphi}
\def\da{\dagger}
\def\om{\omega}
\def\Om{\Omega_{\rm m}}
\def\la{\lambda}
\def\tla{\Lambda}
\def\tr{\tilde{\rho}}
\def\ra{\rightarrow}

\begin{center}
\large{{\bf Is the present expansion of the universe really accelerating?}}\\
\end{center}

\medskip

\begin{center}
\emph{R G Vishwakarma}

\medskip
Inter-University Centre for Astronomy and Astrophysics (IUCAA),\\
Post Bag 4, Ganeshkhind, Pune 411 007, India\\
Email: vishwa@iucaa.ernet.in
\end{center}

\begin{abstract}\noindent
The current observations are usually explained by an accelerating expansion of
the present universe. However, with the present quality of the supernovae Ia data,
the allowed parameter space is wide enough to accommodate the decelerating
models as well. This is shown by considering a particular example of the dark energy
equation-of-state $w_\phi\equiv p_\phi/\rho_\phi=-1/3$, which is equivalent to modifying
the \emph{geometrical curvature} index $k$ of the standard cosmology by shifting  it to
$(k-\alpha)$ where $\alpha$ is a constant.
The resulting decelerating model is consistent with the recent CMB observations
made by WMAP, as well as, with the high redshift supernovae Ia data including SN 1997ff
at $z= 1.755$. It is also consistent with the newly discovered supernovae
SN 2002dc at $z=0.475$ and SN 2002dd at $z=0.95$ which have a general tendency to
improve the fit.

\end{abstract}


\noindent

\noindent
{\bf 1 INTRODUCTION}

\noindent
    It is generally believed that the expansion of the
present universe is accelerating, fuelled by some hypothetical source with
negative pressure collectively known as \emph{'dark energy'}. This belief is
mainly motivated by the high redshift supernovae (SNe) Ia observations, which cannot be
explained by the decelerating Einstein deSitter model which used to be the favoured
model before these observations were made a few years ago. (It may, however, be noted that
if one takes into account the absorption of light by the inter-galactic metallic dust
which extinguishes radiation travelling over long distances, then
the observed faintness of the extra-galactic SNe Ia can be explained successfully
in the framework of the Einstein deSitter model. This issue will be discussed
in a later section. In the rest of the discussion, we shall not include this effect.)

Although the best-fitting standard model to the
SNe Ia data predicts an accelerating expansion, however, it should be noted that the
low density open models, which predict a decelerating expansion, also fit the SNe Ia
observations reasonably well. Unfortunately, these models are ruled out by the recent
measurements of the angular power fluctuations of CMB including the first year observations
made by WMAP.

    In this paper we show, by considering a particular example, that both these observations
(and many others too) can still be explained by a decelerating
model of the universe in the mainstream cosmology.

The observed SNe Ia explosions look fainter than their luminosity in the  Einstein-deSitter
model. This observed faintness is explained by invoking a positive cosmological
constant $\Lambda$ following the fact that the luminosity distance of an
object can be increased by incorporating a \emph{`matter'} with negative
pressure in Einstein's equations.
However, a constant $\Lambda$, is plagued with the so called the cosmological
constant problem: Why don't we see
the large vacuum energy density
$\rho_{\rm v}\equiv \Lambda/8\pi G =\rho_{\rm Pl}\approx 10^{76}$ GeV$^4$,
expected from particle physics which  is $\approx$$10^{123}$ times larger than
the value required by the SNe observations? Or, why are the radiation energy
density and the energy density in  $\Lambda$  ($\Lambda/8\pi G$) set to
an accuracy of better than one part in $10^{123}$ at the Planck time in order
to ensure that the densities in matter and  $\Lambda$ become comparable
at precisely the present epoch?
(Padmanabhan 2002; Peebles \& Ratra 2003; Sahni \& Starobinsky 2000).
A phenomenological solution to understand the smallness of  $\Lambda$ is
supplied by a dynamically decaying $\Lambda$.
A popular candidate is an evolving large-scale scalar field $\phi$, commonly
known as
{\it quintessence}, which does not interact with matter
($T^{ij}_{\phi~ ~;j}=0$) and can
produce negative pressure for a potential energy-dominated field (Peebles \&
Ratra 2003).
This is equivalent to generalizing the equation of state of vacuum to
\be
p_\phi=w_\phi \rho_\phi.
\ee
Moreover, it is always possible to add a term like $T^{ij}_{\phi}$ on the
right hand side of Einstein's field equations independent of the geometry
of the universe.
The equation-of-state parameter $w_\phi$ is, in general, some function of time
and leads to a constant $\Lambda$ for  $w_\phi=-1$.

It would be worthwhile to mention here that there is another approach to
understand the smallness of $\Lambda$ by invoking some phenomenological
models of kinematical  $\Lambda$ which result either from some symmetry
principle (Vishwakarma 2001a, 2001b), or from dimensional analysis (Vishwakarma 2000, 2002a;
Carvalho, Lima \& Waga 1992; Chen \& Wu 1990) or just by assuming $\Lambda$
as a function of the cosmic time $t$ or the scale factor $S(t)$ of the
R-W metric (for a list of this type of models, see Overduin \& Cooperstock 1998).
Although the quintessence fields also behave like a  dynamical $\Lambda$
(with $\Lambda_{\rm effective}\equiv 8\pi G\rho_{\phi}$), they are in
general fundamentally different from the kinematical $\Lambda$ models.
In the former case, quintessence and matter fields are assumed to be conserved
separately, whereas in the
latter case, the conserved quantity is $[T^{ij}_{\rm matter} 
+ \{\Lambda(t)/8\pi G\}g^{ij}]$, which follows from the vanishing divergence 
of the Einstein tensor.
This characteristic of the  kinematical $\Lambda$ models makes them consistent
with Mach's principle (Vishwakarma 2002b). However, in the following, we shall
keep ourselves limited to the former case with a constant $w_\phi$.

\bigskip
\noindent
{\bf 2 FIELD EQUATIONS AND DISTANCE MEASURES}

\noindent
In order to study the constraints on the parameters coming from the different
observations,
we shall describe the model briefly in the following. For the R-W metric,
the Einstein field equations, taken together with conservation of the energy,
yield

\be
H^2(z) = H^2_0 \, \bigg[ \Omega_{{\rm m} 0} \, (1 + z)^3
+ \Omega_{{\rm r} 0}(1 + z)^4 + \Omega_{\phi 0} (1 + z)^{3(1+w_\phi)}
-\Omega_{k0} \, (1 + z)^2 \bigg], \label{eq:hubble}
\ee

\be
q(z) = \frac{H_0^2}{H^2(z)}\left[ \frac{\Omega_{\rm m0}}{2}(1+z)^3+
\Omega_{\rm r0}(1+z)^4 + \frac{(1+3w_\phi)}{2} \Omega_{\phi0}
(1+z)^{3(1+w_\phi)}\right],\label{eq:decel}
\ee
where $\Omega_i$ are, as usual, the different energy density components in
units of the critical density $\rho_{\rm c}\equiv3H^2/8\pi G$:
\be
 \Om \equiv \frac{8 \pi G}{3 H^2} \rho_{\rm m}, \,\hspace{0.25cm}
\Omega_{\rm r} \equiv \frac{8 \pi G}{3 H^2} \rho_{\rm r}, \, ~~
\Omega_{\phi} \equiv \frac{8 \pi G}{3 H^2} \rho_{\phi}, \, ~~
\Omega_k \equiv \frac{k}{S^2 H^2} \label{eq:omegas}
\ee
and the subscript 0 denotes the value of the quantity at the present epoch.
Equation (\ref{eq:hubble}) at $z=0$ implies that
$\Omega_{{\rm m}0}+\Omega_{{\rm r}0}+\Omega_{{\phi}0}=1+\Omega_{k0}$, by the
use of which Equation (\ref{eq:hubble}) reduces to

\newpage
\bq
H^2(z) &=& H^2_0 \, \bigg[ \Omega_{{\rm m} 0} \, (1 + z)^3
+ \Omega_{{\rm r} 0}(1 + z)^4 + \Omega_{\phi 0} (1 + z)^{3(1+w_\phi)}\nonumber \\
&+&(1-\Omega_{{\rm m}0}-\Omega_{{\rm r}0}-\Omega_{{\phi}0}) \, (1 + z)^2 \bigg].
\label{eq:hubblen}
\eq

In a homogeneous and isotropic universe, the different distance measures
of a light source of redshift $z$ located at a radial coordinate distance
$r_1$ are given by the following.

\noindent
The {\it luminosity distance} $d_{\rm L}$:
\be
d_{\rm L} = (1 + z) S_0 ~r_1,
\label{eq:distL}
\ee

\noindent
The {\it angular diameter distance} $d_{\rm A}$:
\be
d_{\rm A}=\frac{S_0 ~r_1}{(1+z)},
\label{eq:distA}
\ee
where $r_1$ is given by
\be
r_1 =
\left\{ \begin{array}{ccl}
\vspace{0.4cm}
\sin\left(\frac{1}{S_0} \, \int_0^z \, \frac{{\rm d} z'}{H(z')} \right),& \mbox{when}& k = 1 \\
\vspace{0.4cm}
\frac{1}{S_0} \, \int_0^z \, \frac{{\rm d} z'}{H(z')}, &\mbox{when}& k = 0 \\
\sinh\left(\frac{1}{S_0} \, \int_0^z \, \frac{{\rm d} z'}{H(z')} \right),& \mbox{when}& k = -1.
\end{array}\right. \label{eq:rdist}
\ee
The present value of the scale factor $S_0$, appearing in
equations (\ref{eq:distL} - \ref{eq:rdist}) which measures the curvature of
spacetime, can be calculated from equation (\ref{eq:omegas}) as
\be
S_0 = H_0^{-1} \sqrt{\frac{k}{
(\Omega_{{\rm m}0}+\Omega_{{\rm r}0}+\Omega_{{\phi}0}-1)}}.\label{eq:szero}
\ee

\bigskip
\noindent
{\bf 3 DEGENERACIES IN FRIEDMANN MODELS COMING FROM SNe Ia OBSERVATIONS}

\noindent
Let us start with the SNe Ia observations which directly measure the apparent
magnitude $m$ and the redshift $z$ of the supernovae.
The theoretically predicted (apparent) magnitude $m(z)$ of a light source
at a given redshift $z$ is given
in terms of its absolute luminosity $M$ and the luminosity distance
$d_{\rm L}$ through the relation
\be
m (z) = \log_{10} [{\cal D}_{\rm L}(z)] + {\cal M}, \label{eq:mageq}
\ee
where  ${\cal D}_{\rm L} \equiv H_0 d_{\rm L}$ is the dimensionless luminosity
distance and ${\cal M} \equiv M - 5 \,\log_{10} H_0 +$ \emph{constant}
(the value of the \emph{constant} depends on the chosen units in which
$d_{\rm L}$ and $H_0$ are measured. For example, if $d_{\rm L}$ is measured
in Mpc and $H_0$ in km s$^{-1}$ Mpc$^{-1}$, then this \emph{constant} comes
out as $\approx$25).

We consider the data
on the redshift and magnitude of a sample of 54 SNe Ia as
considered by Perlmutter et al (excluding 6 outliers from the full
sample of 60 SNe) (Perlmutter et al 1999),
together with SN 1997ff at $z=1.755$ ($m^{\rm eff}=26.02\pm0.34$), the highest
redshift supernova observed so far (Riess et al 2001; Narciso et al 2002).
In addition to this old sample of 55 SNe, we also consider the two newly discovered supernovae
SN 2002dc at $z=0.475$ ($m^{\rm eff}=22.73\pm0.23$) and SN 2002dd at $z=0.95$
($m^{\rm eff}=24.68\pm0.2$), which have been discovered with the help of the
recently installed \emph{Advanced Camera for Surveys} on the
\emph{Hubble Space Telescope} (HST) (Blakeslee et al. 2003).
In the following, we shall also keep an eye on how the addition of
these SNe to the older sample affects the fit.

The $\chi^2$ value is calculated from its usual definition
\be
\chi^2 = \sum_{i = 1}^{N} \,\left[ \frac{m_i^{\rm eff} - m(z_i)}
{\delta m_i^{{\rm eff}}}\right]^2,
\ee
where $N$ is the number of data points (SNe) considered, $m_i^{\rm eff}$
refers to the effective
magnitude of the $i$th SN which has been corrected for the SN lightcurve
width-luminosity relation, galactic extinction and K-correction.
The dispersion
$\delta m_i^{\rm eff}$ is the uncertainty in $m_i^{\rm eff}$.

If one minimizes $\chi^2$ in the standard cosmology ($w_\phi=-1$) by varying
the free parameters $\Omega_{\rm m0}$, $\Omega_{\phi0}=\Omega_{\Lambda0}$ and
${\cal M}$ (the contribution from the term containing $\Omega_{\rm r0}$ is negligible
for this dataset and can be safely neglected), one finds that the best-fitting
solution, from the older sample of 55 SNe,
 is $\Omega_{\rm m0}=0.67$, $\Omega_{\Lambda0}=1.24$ and
${\cal M}=23.92$ with
 $\chi^2=56.93$ at 52 degrees of freedom (dof) ($=$ number of data points $N -$
number of fitted parameters), i.e.,
$\chi^2$/dof $=56.93/52=1.09$, a good fit indeed.
Addition of the new SNe to the older sample gives a similar best-fitting model:
$\Omega_{\rm m0}=0.64$, $\Omega_{\Lambda0}=1.20$ and
${\cal M}=23.92$ with
 $\chi^2$/dof $=57.78/54=1.07$, showing a slight improvement in the fit.

A \emph{`rule of thumb'} for a \emph{moderately} good fit is that $\chi^2$
should be roughly equal to the number of dof. A more quantitative measure
for the \emph{goodness-of-fit} is given by the $\chi^2$-\emph{probability}
which is very often met with in the literature and its compliment is usually
known as the \emph{significance level} (should not be confused with the confidence
regions). If the fitted model
provides a typical value of $\chi^2$ as $x$ at $n$ dof, this probability is
given by
\be
Q(x, n)=\frac{1}{\Gamma (n/2)}\int_{x/2}^\infty e^{-u}u^{n/2-1} {\rm d}u.
\ee
Roughly speaking, it measures the probability that \emph{the model does
describe the data and any discrepancies are mere fluctuations which could have
arisen by chance}. To be more precise, $Q(x, n)$ gives the probability that a model
which does fit the data at $n$ dof, would give a value of $\chi^2$ as large
or larger than $x$. If $Q$ is very small, the apparent discrepancies are
unlikely to be chance fluctuations and the model is ruled out.

The probability $Q$ for the best-fitting standard model comes out as 29.7\%
(without including the new points) and 33.7\% (with the new points),
which represent good fits.
These models represent accelerating expansion of the present universe
according to
equation (\ref{eq:decel}), which  suggests that $q_0<0$ if $w_\phi<-1/3$ and
$\Omega_{\phi0}>\Omega_{\rm m0}/(3\mid w_\phi \mid -1)$.
Thus the best-fitting flat model ($\Omega_{\rm m}+\Omega_{\Lambda}=1$) is
also an accelerating one, which is obtained as
$\Omega_{\rm m0}=1-\Omega_{\Lambda0}=0.31$ with
$\chi^2$/dof = 58.97/53 = 1.11 and $Q=26.6$\%, from the older sample of 55 SNe.
Addition of the new points improves the fit marginally by giving
$\Omega_{\rm m0}=0.32$ with
$\chi^2$/dof = 59.67/55 = 1.08 and $Q=31$\%, as the new best-fitting solution.

Let us now examine the status of the decelerating models, particularly,
the models with
$\Omega_\phi=0$. An important model in this category is the canonical
Einstein-deSitter model ($\Omega_{\rm m0}=1$, $\Omega_{\Lambda0}=0$), which has
a poor fit: $\chi^2/$dof $=93.01/54=1.72$ with $Q=0.08$\%, from the older
sample of 55 points. Even the addition of the new points does not improve the
fit significantly, giving $\chi^2/$dof $=94.61/56=1.69$ with $Q=0.1$\%
and the model is ruled out. However, the open
models with low $\Omega_{\rm m0}$ have reasonable fits. For example, the
following decelerating models with $\Lambda=0$ (even without including the new
points) provide:

\noindent
$\Omega_{\rm m0}=0.2$: $\chi^2$/dof $=66.74/54=1.24$, $Q=11.4$\%;

\noindent
$\Omega_{\rm m0}=0.3$: $\chi^2$/dof $=68.78/54=1.27$, $Q=8.5$\%;

\noindent
$\Omega_{\rm m0}=0.4$: $\chi^2$/dof $=71.31/54=1.32$, $Q=5.7$\%, etc.,

\noindent
which have reasonable fits and by no means are rejectable.
Addition of the new points improves the fits marginally, giving:

\noindent
$\Omega_{\rm m0}=0.2$: $\chi^2$/dof $=67.04/56=1.2$, $Q=14.8$\%;

\noindent
$\Omega_{\rm m0}=0.3$: $\chi^2$/dof $=69.09/56=1.23$, $Q=11.2$\%;

\noindent
$\Omega_{\rm m0}=0.4$: $\chi^2$/dof $=71.69/56=1.28$, $Q=7.7$\%, etc.

\noindent
Open decelerating models
with non-zero $\Lambda$ ($0<\Omega_{\Lambda0}<\Omega_{\rm m0}/2$) have even better fit.
The best-fitting model with a vanishing $\Lambda$ is obtained as
$\Omega_{\rm m0}=0$ with $\chi^2=64.6$ at 54 dof ($\chi^2$/dof $=1.2$) and $Q=15.3$\%,
from the older sample of 55 points. Addition of the new points to this sample
improves the fit by giving $\chi^2$/dof $=65.15/56=1.16$, $Q=18.8$\%.

Figure 1 shows the allowed regions by the full data in the $\Omega_{\rm m0}-{\cal M}$
plane at different confidence levels for a vanishing $\Lambda$ cosmology.
Thus the low $\Omega_{\rm m0}$ open models with a vanishing $\Lambda$ are 
fully consistent
with the current SNe Ia observations. This result is also consistent with the 
findings
of Gott et al (2001) from their median statistics analysis that the open models
with low density and vanishing $\Lambda$ are not inconsistent with the SNe Ia 
data.
Additionally, a low $\Omega_{\rm m0}$
is also consistent with the recent 2DF and Sloan surveys which give
$\Omega_{\rm m0}=0.23\pm0.09$ (Hawkins et al. 2002) and
$\Omega_{\rm m0}\approx 0.14^{+0.11}_{-0.06}$ (at $2\sigma$) (Dodelson et al. 2001).
However, these models do not seem consistent with
the first-year observations of the temperature angular power spectrum of the 
CMB,
measured accurately by NASA's explorer mission ``Wilkinson Microwave Anisotropy Probe" (WMAP)
(Bennett et al., 2003), or even with the earlier CMB observations
(de Bernardis et al. 2000, 2002; Lee et al. 2001; Halverson et al. 2002; Siever et al. 2002).
These observations, when fitted to the standard cosmology,
appear to indicate that $\Omega_{\rm m0}+\Omega_{\Lambda0}\approx 1$.

\bigskip

\begin{figure}[tbh!]
\centerline{{\epsfxsize=14cm {\epsfbox[50 250 550 550]{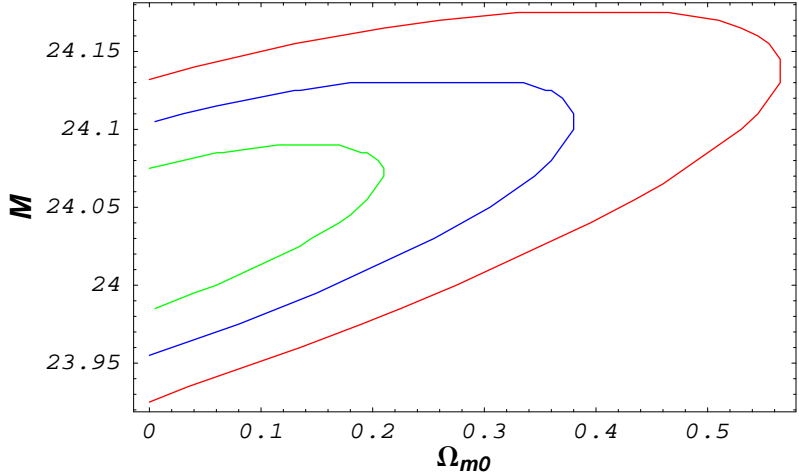}}}}
{\caption{\small The allowed regions by the SNe Ia data (with 57 points) are shown in the
$\Omega_{\rm m0}-{\cal M}$ plane for $\Omega_\phi=0$ cosmology
(${\cal M}$ is usually termed as the `\emph{Hubble constant-free
absolute luminosity}').
The ellipses, in the order of increasing size, correspond to respectively 68\%, 95\% and
99\% confidence levels.}}
 \end{figure}

\bigskip
\noindent
{\bf 4 THE PROPOSED MODEL}

\medskip
\noindent
{\bf 4.1 Motivation}

\noindent
We shall now consider a particular model of dark energy specified by the
equation of state $w_\phi=-1/3$, which, as we shall show in the following, explains both the
observations - SNe Ia, as well as, CMB - very well. The equation of state $w_\phi=-1/3$,
for which $\rho_{\phi}$  varies
as $1/S^2$ (say, $\Omega_{\phi}=\alpha/S^2H^2$),
is interesting in its own right. For this case, the Hubble and the deceleration parameters
reduce respectively to
\be
H^2(z) = H^2_0 \, \bigg[ \Omega_{{\rm m} 0} \, (1 + z)^3
+ \Omega_{{\rm r} 0}(1 + z)^4
+(1-\Omega_{{\rm m}0}-\Omega_{{\rm r}0}) \, (1 + z)^2 \bigg],
\label{eq:hubblef}
\ee
\be
q(z) = \frac{H_0^2}{H^2(z)}\left[ \frac{\Omega_{\rm m0}}{2}(1+z)^3+
\Omega_{\rm r0}(1+z)^4 \right],\label{eq:decelf}
\ee
which describe a decelerating expansion. Interestingly in this case,
$\Omega_{\rm matter}$ ($\equiv \Omega_{\rm m}+\Omega_{\rm r}$) alone is sufficient
to describe $H(z)$ completely, which gives the expansion dynamics of the universe.
The additional knowledge of $\Omega_{{\phi}0}$
decides the curvature through
$\Omega_{k0}=\Omega_{{\rm m}0}+\Omega_{{\rm r}0}+\Omega_{{\phi}0}-1$.
Expressions (\ref{eq:hubblef}) and (\ref{eq:decelf}) are the same as the ones in the
standard cosmology with $\Lambda=0$,
with only one exception: in the standard cosmology,
$\Omega_{{\rm m}0}+\Omega_{{\rm r}0}=1+\Omega_{k0}$, whereas
in the present model,
$\Omega_{{\rm m}0}+\Omega_{{\rm r}0}+\Omega_{{\phi}0}=1+\Omega_{k0}$.
As mentioned earlier, $\Omega_{\phi0}$ contributes to the curvature through
$S_0 = H_0^{-1} \sqrt{k/
(\Omega_{{\rm m}0}+\Omega_{{\rm r}0}+\Omega_{{\phi}0}-1)}$ and hence
to the different distance measures for $k=\pm1$ cases
(for $k=0$, distances do not depend on $S_0$).
Although $\Omega_{\phi0}$ does not contribute to the expansion dynamics of the model
(and neither to the distances for $k=0$ case), it helps $\Omega_{{\rm m}0}$
to assume even those values which are not allowed in the standard cosmology,
and hence gives more leverage to $\Omega_{\rm m0}$. It is in fact this property
of the model which is instrumental
in explaining the CMB observations
even for low  $\Omega_{\rm m0}$, as we shall see in the following.

It may be noted that although the topological defects, like cosmic strings and textures,
also have an equation of state $\rho+3p=0$ (i.e., their density falls off as $1/S^2$),
however, the converse of this is not true and a dark energy with $w_\phi=-1/3$ need not
necessarily be represented by cosmic strings. Moreover, this `pseudo-source' term
does not explicitly contribute to the expansion dynamics (and hence to the deceleration
or to the expansion age) of the model but essentially to the curvature,
as we have mentioned earlier. One can obtain
the same cosmology by removing this term and simultaneously modifying the curvature
index $k$ by ($k-\alpha$) (Vishwakarma \& Singh 2002).
It would, therefore, be more appropriate to consider
it as a shift in the \emph{geometrical curvature} of the standard cosmology
and not as a source term.
This term also appears in the context of brane cosmology by adding a surface term of
brane curvature scalar in the action (Vishwakarma \& Singh 2003;
Singh, Vishwakarma \& Dadhich 2002).

\bigskip
\noindent
{\bf 4.2 CMB Observations}

\noindent
The temperature fluctuations $\Delta T$ in the angular power spectra of the CMB
correspond to oscillations in the $\Delta T$ - $\ell$ (Legendre multipole) space.
The peaks of these oscillations
can be explained in terms of the angle $\theta_{\rm A}$ subtended by the sound horizon at the
last scattering epoch when CMB photons decoupled from baryons at $z_{\rm dec}=1100$
(Hu \& Dodelson 2002). The power spectrum of CMB can be characterized by the positions
of the peaks and ratios of the peak amplitudes. It has long been recognized that
the locations and amplitudes of the peaks in the region $90<\ell<900$ (where the
anisotropies are related to causal processes occurring in the photon-baryon plasma
until recombination) are very sensitive to
the variations in the parameters of the model and hence serve as a sensitive probe
to constrain the cosmological parameters and discriminate among various models
(Hu et al. 2001; Doran \& Lilley 2001, 2002). In fact, for $\ell>40$, the ratios of
the peak amplitudes are insensitive to the intrinsic amplitude of the CMB spectrum.
This renders the positions of
the peaks, particularly the position of the first peak, as a powerful
probe of the parameters of the model.
The first-year observations of WMAP have measured the position of the first peak very
accurately at $\ell=220.1\pm0.8$ (1 $\sigma$) (Page et al. 2003), which we shall
use in our fit.

The angle $\theta_{\rm A}$, which is given by the ratio
of sound horizon to the distance (\emph{angular diameter distance}) of the last
scattering surface, sets the acoustic scale $\ell_{\rm A}$ through
\be
\ell_{\rm A}=\frac{\pi}{\theta_{\rm A}}
=\frac{\pi ~S_0~ r_1 (z_{\rm dec})}{\int_{z_{\rm dec}}^\infty
c_{\rm s}(z) ~ {\rm d} z/H(z)}
,\label{eq:thetaa}
\ee
where  the speed of sound $c_{\rm s}$ in the plasma is given by $c_{\rm s}=1/\sqrt{3(1+R)}$
and $R\equiv3\rho_{\rm b}/4\rho_\gamma=3\Omega_{\rm b0}/[4\Omega_{\gamma0}(1+z)]$
corresponds to the ratio of baryon density to photon density.
The location of $i$-th peak in the angular power spectrum is given by
\be
\ell_{\rm peak_i}=\ell_{\rm A}(i-\delta_i), \label{eq:lpeak}
\ee
where the phase shift $\delta_i$, caused by the plasma driving effect, is
determined predominantly by the pre-recombination physics (Hu et al. 2001) and can
be approximated by
\be
\delta_i = a_i \left\{\frac{\Omega_{\rm r0}(1+z_{\rm dec})}{0.3 ~\Omega_{\rm m0}} \right\}^{0.1},\label{eq:phi}
\ee
where $a_i$ is respectively 0.267, 0.24 and 0.35 for ${\rm peak}_1$, ${\rm peak}_2$
and ${\rm peak}_3$.
Let us recall that $\Omega_{\rm r0}$ gets contributions from photons (CMB) as well as
from neutrinos, i.e.,
$\Omega_{\rm r}=\Omega_{\gamma}+\Omega_{\nu}$. The present photon contribution to the
radiation can be estimated from the CMB temperature $T_0=2.728$K. This gives
$\Omega_{\gamma0}\approx 2.48 ~ h^{-2}\times 10^{-5}$, where $h$ is the present value
of the Hubble parameter in units of 100 km s$^{-1}$ Mpc$^{-1}$.
The neutrino contribution follows from the assumption of 3 neutrino species, a standard
thermal history and a negligible mass compared to its temperature (Hu \& Dodelson 2002)
leading to $\Omega_{\nu0}\approx 1.7 ~ h^{-2}\times 10^{-5}$.

Equations (\ref{eq:thetaa})-(\ref{eq:phi}), suplimented by (\ref{eq:rdist}) and
(\ref{eq:szero}), are now fully
capable to compute the locations of the peaks for given values of the free
parameters $\Omega_{\rm m0}$, $\Omega_{\phi0}$, $\Omega_{\rm b0}$ and $h$.
We notice that a sufficiently big range of these parameters
produce the $\ell_{\rm peak_i}$ values in the observed range. For example,
by fixing $\Omega_{\rm b0}=0.05$ and $h=0.65$, the following choices of $\Omega_{\rm m0}$ and
 $\Omega_{\phi0}$ yield:

\medskip
\noindent
$\Omega_{\rm m0}=0.3$, $\Omega_{\phi0}=0.6$
$\rightarrow$
$\ell_{\rm peak_1}=222.5$,
$\ell_{\rm peak_2}=536.7$,
$\ell_{\rm peak_3}=808.2$;

\medskip
\noindent
$\Omega_{\rm m0}=0.25$, $\Omega_{\phi0}=0.65$
$\rightarrow$
$\ell_{\rm peak_1}=225.0$,
$\ell_{\rm peak_2}=545.1$,
$\ell_{\rm peak_3}=820.9$;

\medskip
\noindent
$\Omega_{\rm m0}=0.2$, $\Omega_{\phi0}=0.72$
$\rightarrow$
$\ell_{\rm peak_1}=220.0$,
$\ell_{\rm peak_2}=540.7$,
$\ell_{\rm peak_3}=814.3$,

\medskip
\noindent
which are in reasonable agreement\footnote{ 
One can have even better fit by adjusting the parameters finely. For example,
one can have

\noindent
$\Omega_{\rm m0}=0.3$, $\Omega_{\phi0}=0.61$
$\rightarrow$
$\ell_{\rm peak_1}=219.8$,
$\ell_{\rm peak_2}=530.2$,
$\ell_{\rm peak_3}=798.4$;

\noindent
$\Omega_{\rm m0}=0.25$, $\Omega_{\phi0}=0.666$
$\rightarrow$
$\ell_{\rm peak_1}=220.3$,
$\ell_{\rm peak_2}=533.6$,
$\ell_{\rm peak_3}=803.5$;

\noindent
with the same $\Omega_{\rm b0}$ and $h$.}
with $\ell_{\rm peak_1}=220.1\pm0.8$ and
$\ell_{\rm peak_2}=546\pm10$ measured by WMAP. 
These measurements are also consistent with
other observations; for example, the BOOMERANG project (de Bernardis et al., 2002)
measured the first three peaks in the ranges:

\noindent
$\ell_{\rm peak_1}:200-223$,
$\ell_{\rm peak_2}:509-561$,
$\ell_{\rm peak_3}:820-857$ at 68\% confidence level; and

\noindent
$\ell_{\rm peak_1}:183-223$,
$\ell_{\rm peak_2}:445-578$,
$\ell_{\rm peak_3}:750-879$ at 95\% confidence level, which are
also in agreement with many other observations like, MAXIMA, DASI, CBI, etc.,
at least on the location of the first peak
(Lee et al. 2001; Halverson et al. 2002;
Sievers et al. 2002).

In Figure 2, we have shown the allowed region in the
$\Omega_{\rm m0}-\Omega_{\phi0}$ plane
which produces the first peak $\ell_{\rm peak_1}=220.1\pm0.8$ (68\% confidence level). The marginalization over the other parameters $\Omega_{\rm b0}$ and 
$h$ 
is achived by taking projection of the full 4-dimensional region on the 
$\Omega_{\rm m0}-\Omega_{\phi0}$ plane (Press et al 1986).

It may be noted that the WMAP results obtained by Spergel et al.(2003)- that only
$w_\phi<-0.78$ can be accommodated within 95\% confidence regions - comes from a
lot of assumptions, some of which are not consistent with many observations. For
example, there are several observations which also measure smaller values of $H_0$,
apart from the higher values (see section 4.4).
However, this degeneracy has not been taken into account and they consider
only that HST observation which gives $H_0=0.72\pm3$(stat)$\pm7$(systematic)
km s$^{-1}$ Mpc$^{-1}$ (Friedman et al. 2001)
close to their best fit value.
Note that there is also another HST Key Project which gives
$H_0=64^{+8}_{-6}$ km s$^{-1}$ Mpc$^{-1}$ (Saurabh et al. 1999).
Sandage
and his collaborators find a value even as low as $H_0=58\pm6$ km s$^{-1}$ 
Mpc$^{-1}$ from an analysis
of SNe Ia distances (Parodi et al. 2000).
In Figure 2, we have instead considered all those possible combinations of
the parameters $\Omega_{\rm m0}\in [0-1]$, $\Omega_{\phi0}\in [0-1.5]$,
$\Omega_{\rm b0}\in[0-0.1]$ and $h\in[0.5-0.8$, which produce
$\ell_{\rm peak_1}=220.1\pm0.8$.
This amounts to a range of $\Omega_{\rm b0} h^2$ as $[0-0.06]$ which safely
contains the range $0.01\leq \Omega_{\rm b0} h^2\leq0.04$ which one essentially
requires to explain the observed abundances of helium, deuterium and lithium
(Narlikar \& Padmanabhan 2001).
It should also be noted that the most favoured values like
$\Omega_{\rm m0}\approx 0.25$ and $\Omega_{\rm b0}h^2\approx 0.02$
(or  $\Omega_{\rm b0}\approx 0.05$) are 
compatible with a moderate $h \approx 0.65$ in this model, as we have 
shown in our examples.

\bigskip
\begin{figure}[tbh!]
\centerline{{\epsfxsize=14cm {\epsfbox[50 250 550 550]{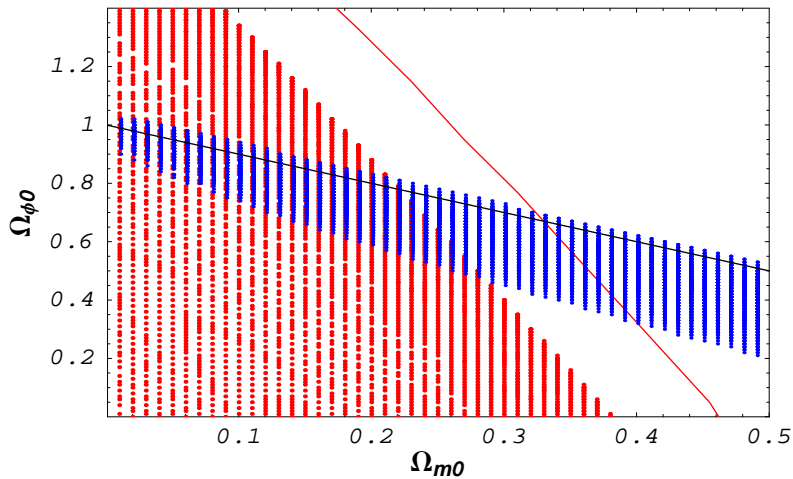}}}}
{\caption{\small The blue region shows the allowed ranges of $\Omega_{\rm m0}$ and
$\Omega_{\phi0}$ by the WMAP which produce the first peak
in the range $220.1\pm0.8$. This has been obtained after marginalizing over
$\Omega_{\rm b0}$ and $h$ (which have been varied in the ranges $[0-0.1]$
and $[0.5-0.8]$ respectively). The red-shaded contour corresponds to 95\% confidence region
(marginalized over ${\cal M}$) from
the SNe Ia data (with all 57 points, including the two newly discovered points
SN 2002dc and SN 2002dd) and the red curve corresponds to the
boundary of the 99\% confidence region given by the same data
(in calculating these regions, the extinction of SNe light by metallic whiskers,
as discussed in section 5, has \emph{not} been taken into account).
The line in black corresponds to the flat model $\Omega_{\rm m0}+\Omega_{\phi0}=1$.}}
\end{figure}

\newpage
\bigskip
\noindent
{\bf 4.3 SNe Ia Observations}

\noindent
For the SNe Ia data in the present model, $\chi^2$ decreases for lower values of
$\Omega_{\rm m0}$ and gives the minimum $\chi^2$ for negative $\Omega_{\rm m0}$.
For example, for the flat model, the minimum value of $\chi^2$ is obtained as
$62.93$ for $\Omega_{\rm m0}=-0.31$
at 53 dof (from the older sample of 55 SNe)
and $63.47$ for $\Omega_{\rm m0}=-0.30$
at 55 dof (with the addition of the new points), which are though not physical.
A `physically viable' best-fitting solution, in this case, can be regarded as
$\Omega_{\rm m0}=0$, which gives
$\chi^2/$dof $=64.56/53=1.22$ for $\Omega_{\phi0}=0.14$ and
${\cal M}=24.03$ with $Q=13.3$\%, from the older sample, which
represents a reasonably good fit, though not as good as the best fit in the
standard cosmology with a constant $\Lambda$. Moreover, the addition of the new points
to this sample improves the fit, giving
$\chi^2/$dof $=65.02/55=1.18$ for $\Omega_{\phi0}=0.23$ and
${\cal M}=24.04$ with $Q=16.7$\%, as the new best-fitting solution.

The allowed region by the data is sufficiently large, as shown in Figure 2, and
low density models, which are also consistent with the WMAP observation, are 
easily accommodated within the
95\% confidence region (Note that the plotted allowed region by WMAP is only at
1$\sigma$ level. At higher levels, the region will be wider).
For example, the following models represent reasonable fit.

\noindent
$\Omega_{\rm m0}=0.3$, $\Omega_{\phi0}=0.6$:
$\chi^2/$dof $=71.93/54=1.33$ with $Q=5.2$\%;

\noindent
$\Omega_{\rm m0}=0.25=1-\Omega_{\phi0}$:
$\chi^2/$dof $=71.36/54=1.32$ with $Q=5.7$\%;

\noindent
$\Omega_{\rm m0}=0.25$, $\Omega_{\phi0}=0.65$:
$\chi^2/$dof $=70.76/54=1.31$ with $Q=$6.3\%;

\noindent
$\Omega_{\rm m0}=0.2$, $\Omega_{\phi0}=0.7$:
$\chi^2/$dof $=69.63/54=1.29$ with $Q=7.5$\%, etc., which are obtained from
the older sample of 55 SNe.
Addition of the new points improves the fits, giving:

\noindent
$\Omega_{\rm m0}=0.3$, $\Omega_{\phi0}=0.6$:
$\chi^2/$dof $=72.44/56=1.29$ with $Q=6.9$\%;

\noindent
$\Omega_{\rm m0}=0.25=1-\Omega_{\phi0}$:
$\chi^2/$dof $=71.87/56=1.28$ with $Q=7.5$\%;

\noindent
$\Omega_{\rm m0}=0.25$, $\Omega_{\phi0}=0.65$:
$\chi^2/$dof $=71.22/56=1.27$ with $Q=$8.3\%;

\noindent
$\Omega_{\rm m0}=0.2$, $\Omega_{\phi0}=0.7$:
$\chi^2/$dof $=70.06/56=1.25$ with $Q=9.8$\%, etc.

\noindent
One can go up to even as high as $\Omega_{\rm m0}\approx 0.4$ at 99\% 
confidence level.

\bigskip
\noindent
{\bf 4.4 Age of the Universe}

\noindent
The parameters $H_0$ and $\Omega_{\rm m0}$ set the age of the universe in this
model. Remember that there is still quite large uncertainty in the present value of $H_0$. Sandage
and his collaborators find $H_0=58\pm6$ km s$^{-1}$ Mpc$^{-1}$ from an analysis
of SNe Ia distances (Parodi et al. 2000). This is also consistent with the value
obtained from an analysis of clusters using Sunyaev-Zeldovich effect which gives
$H_0=60\pm10$ km s$^{-1}$ Mpc$^{-1}$ (Birkinshaw 1999). An HST
Key Project supplies $H_0=64^{+8}_{-6}$ km s$^{-1}$ Mpc$^{-1}$ (Saurabh et al. 1999).
Some experiments also measure higher $H_0$, for example, another HST Key Project,
which uses Cepheids to calibrate several different secondary distance indicators,
finds $H_0=72\pm3$(stat)$\pm7$(systematic) km s$^{-1}$ Mpc$^{-1}$ (Friedman et al. 2001).
Also by using the fluctuations
in the surface brightness of galaxies as distance measure, it has been found that
$H_0=74\pm4$ km s$^{-1}$ Mpc$^{-1}$ (Blakeslee et al. 1999). Thus it appears that
$H_0$ lies somewhere in the range (50 $-$ 78) km s$^{-1}$ Mpc$^{-1}$.
An average value of $H_0=65$ km s$^{-1}$ Mpc$^{-1}$ from this range, constrains
$\Omega_{\rm m0}$ of the model by $0<\Omega_{\rm m0}\leq 0.34$
to give the age of the universe $t_0 \geq 12$ Gyr, so that the age of the
oldest objects detected so far, e.g., the globular clusters of age $t_{\rm GC}=12.5
\pm 1.2$ Gyr (Cayrel et al. 2001; Gnedin et al. 2001), can be explained.
The presently favoured value $\Omega_{\rm m0}\approx$$0.25$ easily accommodates in this range.
It is interesting to note that the average value of 
$H_0=65$ km s$^{-1}$ Mpc$^{-1}$ we have preferred, is in good agreement with 
$H_0=67\pm2$(stat)$\pm5$(systematic) km s$^{-1}$ Mpc$^{-1}$ obtained recently
by Gott et al (2001) from an analysis based on the median statistics.

\bigskip
\noindent
{\bf 5 EXTINCTION BY METALLIC DUST}

\noindent
In this section, we shall discuss the absorption of light by metallic dust 
ejected from the SNe explosions $-$ an issue
which is generally avoided while discussing $m$-$z$ relation
for SNe Ia. 
Although a number of observers believe this to be not a significant effect
(see, for example, Riess et al (2001)), however, taking this effect into 
consideration does improve the fit to the data, as we shall see in the 
following.

It is well known that the metallic vapours are ejected from the SNe
explosions which are subsequently pushed out of the galaxy through pressure
of shock waves (Hoyle \& Wickramasinghe, 1988; Narlikar et al, 1997).
Experiments show that metallic vapours on cooling, condense into elongated
whiskers of $\approx$$0.5-1$ mm length and $\approx$$10^{-6}$ cm 
cross-sectional radius (Hoyle et al, 2000).
Indeed this type of dust extinguishes radiation travelling over long distances
(Aguire, 1999; Vishwakarma, 2002a). The density of the dust can be estimated
along the lines of Hoyle et al (2000). If the metallic whisker production is
taken as 0.1 $M_\odot$ per SN and if the SN production rate is taken as
1 per 30 years per galaxy, the total production per galaxy (of spatial density
$\approx$ 1 per $10^{75}$ cm$^3$) in $10^{10}$ years is $\approx 2/3\times 
10^{41}$ g. The expected whisker density, hence, becomes 
$2/3\times 10^{41}\times 10^{-75}\approx 10^{-34}$ g cm$^{-3}$.
We shall later see that this value is in striking agreement with the
best-fitting value coming from the SNe Ia data.

In an isotropic and homogeneous universe, the contribution to the effective magnitude
arising from the
absorption of light by the intervening whisker-like dust, is given by

\begin{equation}
\Delta m(z)=\int_0^{l(z)} \kappa \rho_g {\rm d}l=\kappa \rho_{g0}\int_0^z
(1+z')^2\frac{{\rm d}z'}{H(z')},
\end{equation}
where $\kappa$ is the mass absorption coefficient, which is effectively
constant over a wide range of wavelengths and is of the order $10^5$
cm$^2$ g$^{-1}$ (Wickramasinghe \& Wallis, 1996),
$\rho_g\propto S^{-3}$ is the whisker grain density and $l(z)$ is the 
proper distance traversed by light through
the inter-galactic medium emitted at the epoch of redshift $z$.
The net magnitude is then given by

\begin{equation}
 m^{\rm net}(z)=m(z) + \Delta m(z),
\end{equation}
where the first term on the r.h.s. corresponds to the usual magnitude from the
cosmological evolution given by equation (\ref{eq:mageq}).

We note that taking account of this effect improves the fit to the SNe Ia data considerably.
For example, the model with $\Omega_{\rm m0}=0.25$, $\Omega_{\phi0}=0.65$ now gives
$\chi^2/$dof $=65.25/53=1.23$ for ${\cal M}=24.03$ and $\rho_{g0}=2.02\times 
10^{-34}$ g cm$^{-3}$
with $Q=12.1$\%, from the older sample of 55 points.
The addition of the new points improves the fit further by giving
$\chi^2/$dof $=65.75/55=1.20$ for ${\cal M}=24.03$ and $\rho_{g0}=1.89\times 
10^{-34}$ g cm$^{-3}$
with $Q=15.2$\%.
Even the Einstein-deSitter model
($\Omega_{\rm m0}=1$, $\Omega_{\phi0}=0$) gives an acceptable fit:
$\chi^2/$dof $=68.40/53=1.29$ for ${\cal M}=24.04$ and $\rho_{g0}=4.95\times 
10^{-34}$ g cm$^{-3}$
with $Q=7.6$\% (from the 55 points-data) and
$\chi^2/$dof $=68.97/55=1.25$ for ${\cal M}=24.04$ and $\rho_{g0}=4.75\times 
10^{-34}$ g cm$^{-3}$
with $Q=9.8$\% (by adding the new points).
Interestingly, this model (Einstein-deSitter) is also
consistent with the CMB observations: $\Omega_{\rm b0}=0.05$ and $h=0.65$ yield
$\ell_{\rm peak_1}=202.9$,
$\ell_{\rm peak_2}=476.8$,
$\ell_{\rm peak_3}=717.6$. One can improve the fit by increasing
$\Omega_{\rm b0}$ and/or decreasing $h$: $\Omega_{\rm b0}=0.1$ and $h=0.55$ 
yield
$\ell_{\rm peak_1}=220.4$,
$\ell_{\rm peak_2}=521.4$,
$\ell_{\rm peak_3}=784.9$. 
Also a better fit can be obtained in
open models. For example, the model $\Omega_{\rm m0}=0.87$,
$\Omega_{\phi0}=0$ with $\Omega_{\rm b0}=0.05$ and $h=0.65$ yields
$\ell_{\rm peak_1}=220.1$,
$\ell_{\rm peak_2}=518.8$,
$\ell_{\rm peak_3}=780.9$. This model also has an acceptable fit to the SNe Ia data:
$\chi^2/$dof $=67.89/53=1.28$ for ${\cal M}=24.04$ and $\rho_{g0}=4.34\times 
10^{-34}$ g cm$^{-3}$
with $Q=8.2$\% (from the 55 points-data) and
$\chi^2/$dof $=68.45/55=1.24$ for ${\cal M}=24.04$ and $\rho_{g0}=4.15\times 
10^{-34}$ g cm$^{-3}$ with
$Q=10.5$\% (by adding the new points).
However, these models suffer from the age problem
if $h$ is not sufficiently low. For example, $h$ should be $\leq 54$ for $\Omega_{\rm m0}=1$
with $\Omega_{\phi0}=0$.
Additionally, there is much evidence for low  $\Omega_{\rm m0}$, as reviewed
by Peebles \& Ratra (2003).

\bigskip
\noindent
{\bf 6 CONCLUSION}

\noindent
    There seems to be an impression in the community that the current observations,
particularly the high redshift SNe Ia observations and the measurements of the angular
power fluctuations of the CMB, can be explained only in the framework of
an accelerating
universe. This, however, does not seem correct. The allowed parameter space by the
datasets is wide enough to accommodate decelerating models also.
We have shown that both these observations can also be explained in a decelerating
low density-model with a dark energy equation-of-state $w_\phi=-1/3$ and the preferred
curvature of the spatial section is slightly negative.
For this equation of state, the resulting `dark energy' does not contribute to
the expansion dynamics of the model (described by the Hubble parameter) and contributes
to the curvature only.

In order to fit the model to the data, we have considered the most recent observations.
For example, for the SNe data, we have considered the older sample of 55 SNe
of type Ia (54 SNe
used by Perlmutter et al $+$ SN 1997ff at $z=1.755$) together with the two
newly discovered supernovae SN 2002dc at $z=0.475$ and SN 2002dd at $z=0.95$.
Addition of these new points to the older sample improves the fit to,
more or less, all  the models. For CMB, we have considered the first-year observations
of WMAP which have measured the position of the first peak very
accurately.

It may be noted that the case $w_\phi=-1/3$ is
not special in any sense to the datasets and slightly more decelerating
or slightly less decelerating models are also consistent with both the observations.
In order to verify this, we change $w_\phi$ slightly from $w_\phi=-1/3$ (in both
directions) to, say, $w_\phi=-0.3$ and $w_\phi=-0.35$ and check the status of
the resulting models. The result is the following.

\medskip
\noindent
$w_\phi=-0.3$:

\noindent
A test model, for example,  $\Omega_{\rm m0}=0.25$, $\Omega_{\phi0}=0.65$,
$\Omega_{\rm b0}=0.05$ and $h=0.65$ yields
$\ell_{\rm peak_1}=220.2$,
$\ell_{\rm peak_2}=533.4$,
$\ell_{\rm peak_3}=803.2$. The SNe Ia data, for the same model, give
$\chi^2/$dof $=72.10/54=1.34$ with $Q=5.0$\%, from the older sample of 55 points
and $\chi^2/$dof $=72.60/56=1.30$ with $Q=6.7$\%, by adding the new points to this sample.

\medskip
\noindent
$w_\phi=-0.35$:

\noindent
The same test model, in this case, yields
$\ell_{\rm peak_1}=227.3$,
$\ell_{\rm peak_2}=550.5$,
$\ell_{\rm peak_3}=829.1$.
The SNe Ia data, for this case, give
$\chi^2/$dof $=70.12/54=1.3$ with $Q=6.9$\% , from the older sample of 55 points
and $\chi^2/$dof $=72.57/56=1.26$ with $Q=9.1$\%, by adding the new points to
this sample.
These fits are acceptable
and comparable to their respective values for $w_\phi=-1/3$ mentioned in sections 4.2
and 4.3.

We also note that if  we take into account the extinction of SNe light by
the inter-galactic metallic dust, then the observed dimming of the high redshift
SNe Ia can also be explained
by the models without any dark energy, such as, the Einstein-deSitter model.
These models are also consistent with CMB observations. In fact, there is
 a degeneracy in the $\Omega_{\rm m0}-\Omega_{\phi0}$ plane along a line
$\Omega_{\rm m0}+\Omega_{\phi0}\approx 1$ and a wide range of $\Omega_{\rm m0}$
is consistent with the  WMAP observation.

Interestingly, another alternative explanation of the observed faintness of SNe Ia at
large distances can be given in terms of a quantum mechanical oscillation between
the photon field and a hypothetical axion field in the presence of extra-galactic
magnetic fields. To satisfy other cosmological constraints, one then simply needs
some form of uniform dark energy with $w_\phi\approx-1/3$ and the universe would be
decelerating (Csaki et al. 2002).

We conclude that it is premature to claim, on the basis of the existing data,
that the present expansion of the universe is accelerating (or decelerating).
Only more accurate SNe Ia data with $z$ significantly $>1$ can remove this ambiguity,
as $q$ is sensitive
significantly to the SNe Ia data only. Whereas the CMB observations are consistent with both -
accelerating as well as decelerating models, as mentioned above.
This endeavour may be accomplished by
the proposed {\it SuperNova Acceleration Probe} (SNAP) experiment
which aims to give accurate luminosity distances
of type Ia SNe up to $z\approx 1.7$.

\medskip
\noindent
{\bf ACKNOWLEDGEMENTS}

\noindent
The author thanks DAE for his Homi Bhabha postdoctoral fellowship and
T. Padmanabhan for useful comments and discussions.

\bigskip
\noindent
{\bf REFERENCES}\\
Aguire A. N., 1999, ApJ, 512, L19\\
Bennett et al., astro-ph/0302207\\
de Bernardis P., et al., 2002, ApJ., 564, 559\\
Blakeslee J. P., et al., 1999, ApJ. Lett., 527, 73\\
Blakeslee J. P., et al., astro-ph/0302402\\
Birkinshaw M., 1999, Phys. Rep., 310, 97\\
Cayrel R., et al, 2001, Nature, 409, 691\\
Carvalho J. C., Lima J. A. S., Waga I., 1992, Phys. Rev. D, 46, 2404\\
Chen W., Wu Y. S., 1990, Phys. Rev. D, 41, 695\\
Csaki C., Kaloper N., Terning J., 2002, Phys. Rev. Lett., 88, 161302\\
Dodelson S., et al., astro-ph/0107421\\
Doran M., Lilley M., Schwindt J., Wetterich C., 2001, ApJ., 559, 501\\
Doran M., Lilley M., 2002, MNRAS, 330, 965\\
Freedman W. L. et al., 2001, ApJ., 553, 47\\
Gnedin O. Y., Lahav O., Rees M. J., astro-ph/0108034\\
Gott III J. R., Vogeley M. S., Podariu S., Ratra B., 2001, ApJ, 549, 1\\
Halverson N. W. et al., 2002, ApJ., 568, 38\\
Hawkins E., et al., astro-ph/0212375\\
Hoyle F., Wickramasinghe N. C., 1988, Astrophys. Space Sc. 147, 245\\
Hoyle F., Burbidge G., Narlikar J. V., 2000, {\it A Different Approach to

\hspace{.5cm}       Cosmology}, (Cambridge: Cambridge Univ. Press)

\noindent
Hu W., Dodelson S., 2003, Ann. Rev. Astron. Astrophys., 40, 171\\
Hu W., Fukugita M., Zaldarriaga M., Tegmark M., 2001, ApJ., 549, 669\\
Lee A. T. et al., 2001, ApJ., 561, L1 \\
Narciso B., et al., 2002, ApJ., 577, L1 (astro-ph/0207097)\\
Narlikar J. V., Wickramasinghe N. C., Sachs R., Hoyle F., 1997, Int. J. Mod.

\hspace{.5cm}     Phys. D, 6, 125

\noindent
Narlikar J. V., Padmanabhan T., 2001, Annu. Rev. Astron. Astrophys., 39, 

\hspace{.5cm} 211

\noindent
Overduin J. M., Cooperstock F. I., 1998, Phys. Rev. D, 58, 043506\\
Padmanabhan T., hep-th/0212290\\
Page et al., astro-ph/0302220\\
Parodi B. R., et al., 2000, ApJ., 540, 634\\
Peebles P. J. E., Ratra B., 2003, Rev. Mod. Phys., 75, 559 (astr0-ph/0207347)\\
Perlmutter S., et al., 1999, ApJ., 517, 565\\
Press W. H., Teukolsky S. A., Vetterling W. T., Flannery B. P., 1986, 

\hspace{.5cm} {\it Numerical Recipes}, (Cambridge University Press) 

\noindent
Riess A. G., et al., 2001, ApJ., 560, 49\\
Sahni V., Starobinsky A., 2000, Int. J. Mod. Phys. D 9, 373\\
Saurabh J., et al., 1999, ApJ. Suppl., 125, 73\\
Sievers J. L. et al., astro-ph/0205387\\
Singh P., Vishwakarma R. G., Dadhich N., hep-th/0206193\\
Spergel D. N. et al., astro-ph/0302209\\
Vishwakarma R. G.,  2000, Class. Quantum Grav., 17, 3833\\
Vishwakarma R. G.,  2001a, Gen. Relativ. Grav., 33, 1973\\
Vishwakarma R. G.,  2001b, Class. Quantum Grav., 18, 1159\\
Vishwakarma R. G.,  2002a, MNRAS, 331, 776\\
Vishwakarma R. G.,  2002b, Class. Quantum Grav., 19, 4747\\
Vishwakarma R. G., Singh P., 2003, Class. Quantum Grav., 20, 2033\\
Wickramasinghe N. C., Wallis D. H., 1996, Astrophys. Space Sc.

\hspace{.5cm} 240, 157

\end{document}